# Anisotropic Electrene T′-Ca$_2$P with Electron Gas Magnetic Coupling as Anode Material for Na/K Ion Batteries

Jiaxin Jiang,[a,†] Kai Wang,[a,†] Hongyan Guo,[a] Guizhong Zuo,[b] Zhiwen Zhuo,[*a] and Ning Lu[*a]

[a]Anhui Province Key Laboratory of Optoelectric Materials Science and Technology, Key Laboratory of Functional Molecular Solids Ministry of Education, and Department of Physics, Anhui Normal University, Wuhu, Anhui 241000, China

[b]Institute of plasma physics, HIPS, Chinese academy of Sciences, Hefei, 230031, China

*Corresponding authors: Zhiwen Zhuo, E-mail: zhuozw@ahnu.edu.cn, Anhui Normal University, Wuhu, Anhui 241000, China

Ning Lu, E-mail: luning@ahnu.edu.cn, Anhui Normal University, Wuhu, Anhui 241000, China

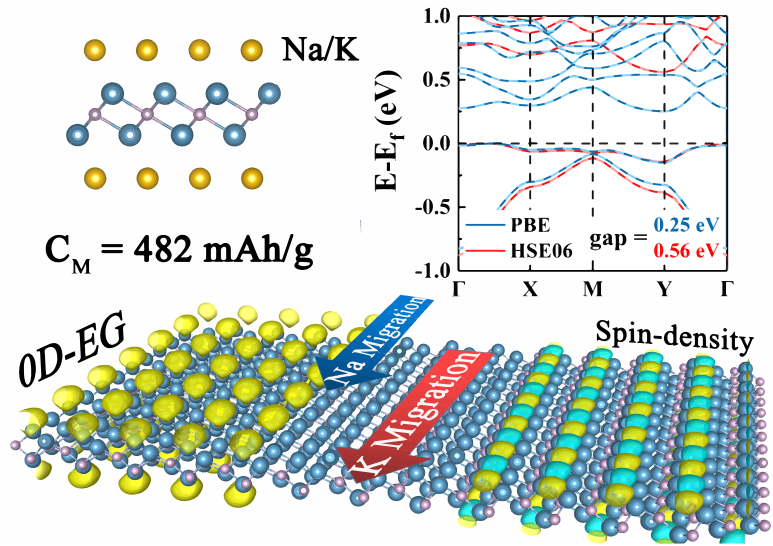

## ABSTRACT

There is an urgently need for the high-performance rechargeable electrical storage devices as supplement or substitutions of lithium ion batteries due to the shortage of lithium in nature. Herein we propose a stable 2D electrene T′-$Ca_2P$ as anode material for Na/K ion batteries by first-principle calculations. Our calculated results show that T′-$Ca_2P$ monolayer is an antiferromagnetic semiconducting electrene with spin-polarized electron gas. It exhibits suitable adsorption for both Na and K atoms, and its anisotropic migration energy barriers are 0.050/0.101 eV and 0.037/0.091 eV in $b/a$ direction, respectively. The theoretical capacities for Na and K are both 482 mAh/g, while the average working voltage platforms are 0.171-0.226 V and 0.013-0.267 V, respectively. All the results reveal that the T′-$Ca_2P$ monolayer has promised application prospects as anode materials for Na/K ion batteries.

# 1. Introduction

Lithium-ion batteries as the most successful clean energy storage devices are widely used in many fields, e.g., electronic devices, vehicles[1], satellites[2], power stations, etc.[3] However, lithium resources on earth are relatively rare and predictably depleted[4], it is necessary to design other alternative rechargeable batteries. Recently, Na-ion batteries[5] (SIBs) and K-ion batteries[6] (PIBs) have drawn increasing attention not only for the large number of resources reserves as well as associated low cost, but also the similar electrochemical mechanisms to LIBs. However, SIBs and PIBs are still in the initial development stage. It is particularly important to design and research electrode materials with desirable properties for the development of SIBs and PIBs[7]. As one key issue for SIBs and PIBs, searching high-performance anode materials is still a great challenge. One of the main reasons is that the naturally large size of Na[8] and K atoms hinder the migration and intercalation of ion[9], many high-performance anode materials in LIBs are not applicable to SIBs and PIBs. Graphite[10], as a commercial anode material for LIBs, was considered to be the promising anode material for SIBs and PIBs due to its unique layered structure and good electrical conductivity. However, conventional graphite is not a suitable anode material for SIBs[11] and the de-intercalation of K is also accompanied by problems such as large volume deformation and structural degradation[12]. Though improved graphite-based materials such as CNC, expanded graphite, and hard carbon have been developed[13-17], the cycle capacity is still less than 300 mAh/g. Alloy electrode materials have higher theoretical adsorption capacities, such as Sn-based materials[18] (500 mAh/g for SIBs), Sb-based materials[19] (660 mAh/g for PIBs), but the severe volume expansion become a huge impact on the stability and safety of the electrode. Thus, the exploration of promising anode materials for SIBs and PIBs is still urgent.

Electride materials[20] are a special kind of ionic compounds contains intriguing confined anionic electrons. The diverse distributions of interstitial electrons in electrides, leading four topologies of electron gas[21], 0D cavities[22], 1D-linked[23] channels (1DEG), 2D planes[24] (2DEG) and 3D configurations[25], endowed unique properties for electrides, render them wide applications in catalysis[26], rechargeable batteries[27], magnetic materials[28], and superconductors[29]. Their 2D structure, known as electrenes, are layered electride materials generally with electron gas formed by free-electron confined on the both surfaces, leading a very high electron mobility, ultra-low work function, and wide application in electrode material to achieve high specific capacity ($C_M$), low average open circuit voltage ($V_{OC}$), and low ion migration barrier energy ($E_b$). For instance, the $Ca_2N$ monolayer[30] was predicted as anode materials for SIBs[31] with theoretical $C_M$ of 1138 mAh/g, low migration $E_b$ of 0.084 eV and low $V_{OC}$ of 0.18 V, which was confirmed by experimentally synthesized $Ca_2N$ electride[27] with observed initial $C_M$ of 1110 mAh/g and cycle $C_M$ of 320 mAh/g as anode for SIBs. $Y_2C$[32-33] exhibits a theoretical $C_M$ of 564 mAh/g, a lower migration $E_b$ of 0.01 eV and low $V_{OC}$ of 0.24-0.45 V. $M_2X$ compounds[34] (M = group II-A metal, X = C, N, P, As and Sb), and concluded that $Sr_2N$, $Ba_2N$, $Sr_2P$, $Ba_2As$, $Ba_2P$ and $Ba_2Sb$ are dynamically stable as well as promising to be anode for SIBs and PIBs with low migration energy barriers in range of only 5-32 meV.

Here, we propose a 2D anisotropic $Ca_2P$ phase, T′-$Ca_2P$ monolayer, reconstructed from unstable T-$Ca_2P$ phase[34] predicted in previous work as anode for SIBs and PIBs based on density functional theory (DFT). The dynamical and thermal stabilities of T′-$Ca_2P$ monolayer are confirmed by the computation of phonon dispersion and ab initio molecular dynamics (AIMD) simulations. Remarkably, the ground state of T′-$Ca_2P$ monolayer is novel antiferromagnetic (AFM) semiconductor with 0D electron gas as well as 0D spin density on its both surfaces. The calculated results show that both Na and K atom has a well adsorption on T′-$Ca_2P$ surface as well as a low migration $E_b$ of 0.050 eV and 0.037 eV along *b*-direction, respectively. The theoretical $C_M$ for SIBs and PIBs are both 482 mAh/g with low $V_{OC}$ of 0.171-0.226 V and 0.013-0.267 V, respectively. These findings suggest that T′-$Ca_2P$ is promising candidates for anode material for SIBs and PIBs.

## 2. Computational details

All the calculations were performed using spin-polarized density functional theory (DFT) in conjunction with the projector-augmented wave[35-36] (PAW) scheme and performed in the Vienna ab initio Simulation Package[36-37] (VASP). The exchange correlation energy is treated using the generalized gradient approximation (GGA) as formulated by Perdew-Burke-Ernzerhof[38] (PBE), and the DFT-D2 method of Grime[39] is adopted to describe the vdW interactions between the layers. The cutoff energy for the plane-wave basis expansion is set to 500 eV. A Γ-center $6 \times 10 \times 1$ k-points mesh via Monkhorst-Pack method is used for T′-Ca$_2$P geometry optimization, while $12 \times 20 \times 1$ is used for the static total energy calculation. The self-consistent convergence of total energy and force is set to $10^{-7}/10^{-5}$ eV and 0.001/0.01 eV/Å for unit cell and supercell, respectively. To avoid any interactions between adjacent periodic images, a vacuum separation of 20 Å is set along the *c* direction of the monolayer structures. An ab initio molecular dynamics (AIMD) simulations are performed to assess the thermal stability of T′-Ca$_2$P monolayer at 300 K, 600 K, 900 K, 1200 K and 1500 K within time step of 1 fs and total time of 5 ps. The migration energy barriers are calculated based on the climbing image nudged elastic band (CI-NEB) method. VASPKIT code[40] were used for calculation post-processing.

## 3. Results and discussion

### 3.1 Structure, stability, electronic and mechanical properties of T′-Ca$_2$P monolayer

The geometric structure of Ca$_2$N electrene-like T-Ca$_2$P structure is shown in Fig. 1a. The fully optimized T′-Ca$_2$P structure after phase-transition from unstable T-Ca$_2$P is lightly puckered with anisotropic atom arrangement (Fig. 1a). The lattice and total energy of the nonmagnetic (NM), ferromagnetic (FM), and three different antiferromagnetic (AFM) states (AFM-0D, AFM-1D$_1$ and AFM-1D$_2$) of T′-Ca$_2$P monolayer are considered and calculated as shown in Fig. 1b. The energy of all magnetic states is lower than that of NM state by 31 to 49 meV per formula (Ca$_2$P, meV/f.), of which the AFM-0D state exhibits lowest energy in all states. The energy of FM state slightly lower the other AFM states by 1 meV/f. for AFM-1D$_1$ and 2 meV/f. for AFM-1D$_2$, higher than AFM-0D state by 16 meV/f.. The space group of the T′-Ca$_2$P monolayer is $P2_1/m$ and its unit cell consists of four Ca and two P atoms. The lattice constants of FM sate are predicted to be a = 7.38 Å, b = 4.23 Å with three Ca-P bond lengths of 2.82, 2.83 and 2.85 Å and thickness d$_Z$ of 3.00 Å, respectively. The lattice constants of AFM-0D slightly change into a = 7.35 Å, b = 8.41 Å and thickness d$_Z$ of 3.07 Å (see part I for more structural information of T-Ca$_2$P and T′-Ca$_2$P in the ESI†).

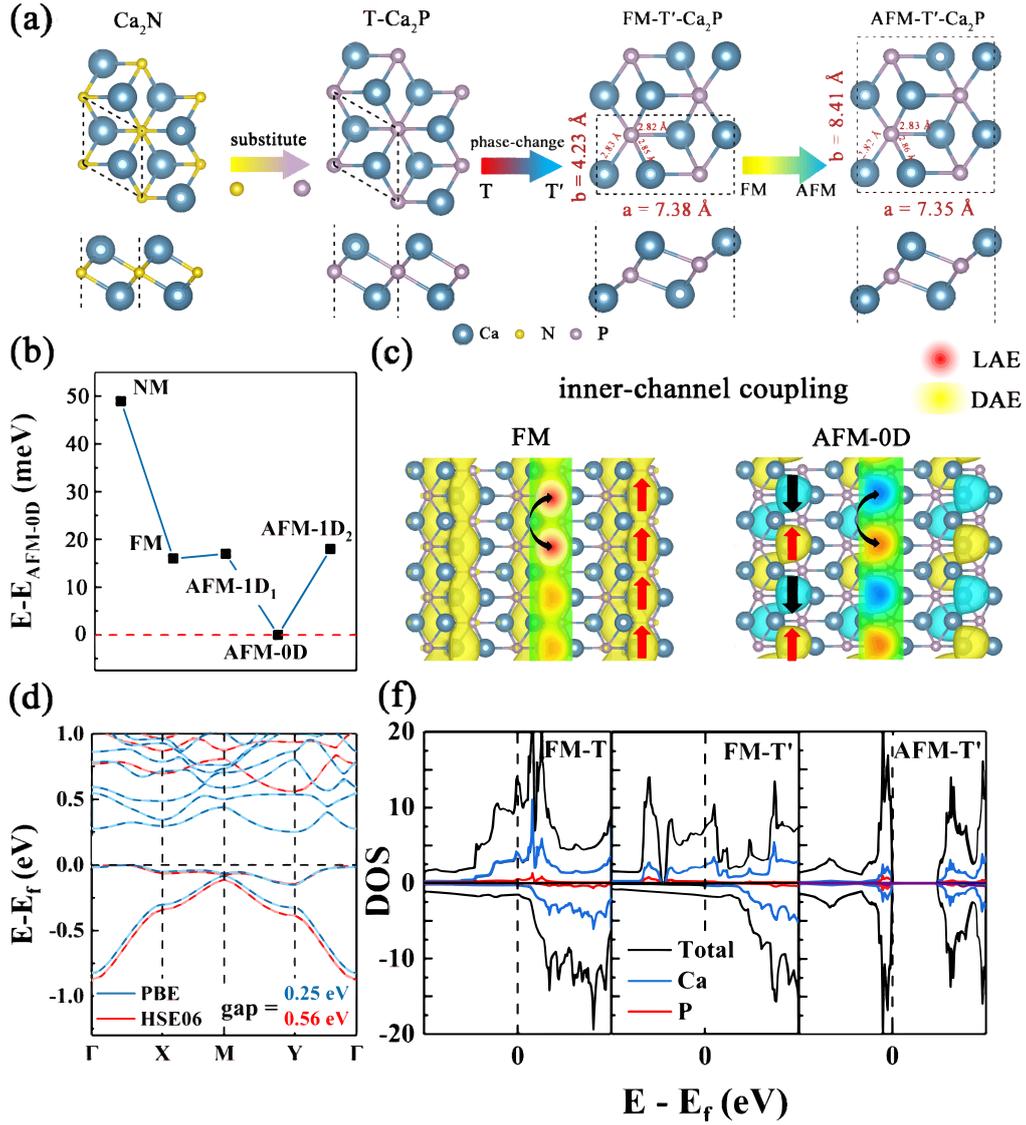

**Fig. 1** (a) The top and side view of optimized $Ca_2N$ $2 \times 2$ supercell, $T$-$Ca_2P$ $2 \times 2$ supercell, $T'$-$Ca_2P$ (FM) $1 \times 2$ supercell and $T'$-$Ca_2P$ (AFM) unit-cell, respectively. The dashed lines indicate a unit cell. (b) Calculated energy difference between NM, FM, AFM-$1D_1$ and AFM-$1D_2$ states of $T'$-$Ca_2P$ monolayer with AFM-0D states of $T'$-$Ca_2P$, respectively. (c) The spin density of the FM and AFM-0D states of $T'$-$Ca_2P$. The isovalue is set at 0.002 e/Bohr$^3$. (d) Calculated spin-polarized band structure of $T'$-$Ca_2P$ monolayer. The black dashed line represents the Fermi-level. (f) Calculated total and partial DOS of FM state of $T$-$Ca_2P$ and FM, AFM states of $T'$-$Ca_2P$.

The magnetism of T′-Ca$_2$P is mainly contributed from the extra electron (e$_{ex}$) on the surface of (Ca$_2$P)$^+$ monolayer (under the consideration of T′-Ca$_2$P = [(Ca$_2$P)$^+$(e$_{ex}$)$^-$]), which distributed on the centers of three Ca atoms on the same surface (See Fig. S2 in ESI†). The spin charge density of FM state is linearly distributed between Ca atoms but away from their centers, as parallel 1D channel-like spin density on its both surfaces. The magnetic moment of FM state is calculated to be 1.32 $\mu_B$ per unit cell (Ca$_4$P$_2$), equaling each extra electron contributes 0.66 $\mu_B$ with a high spin-polarization degree of 66%. The AFM-1D$_1$ and AFM-1D$_2$ states show similar spin charge density with FM, while AFM-0D state shows 0D spin density with staggered distributions of spin-up and spin-down density in the channel. According to the theory proposed by Huang et al.[41], the coexistence of localized anionic electrons (LAE) and delocalized anionic electrons (DAE), and nearly half-filled LAE leading to the Stoner's-type instability. The spin splitting results in the magnetic moment[42] on LAE distributed over the T′-Ca$_2$P monolayer surface. The calculated energy difference between different magnetic states mentioned above indicates that the direct interaction of inner-channel is obviously stronger than the interaction of inter-channel. The reason may originate from that the relatively strong $d$(Ca)-$d$(Ca) direct exchange due to the close distance of neighbor LAEs in each channel (see part II for more details of magnetic coupling in ESI†).

The lattice dynamical and thermal stabilities of T′-Ca$_2$P monolayer are evaluated by the phonon dispersion curves and ab initio molecular dynamics (AIMD) simulations. As shown in the Fig. S3a and S3b†, there were no imaginary frequency in the whole Brillouin zone, indicated the high dynamical stability of the FM and AFM-0D states of T′-Ca$_2$P. The inset of Fig. S3c-S3g† are the snapshots of structures after the simulations of 5.0 ps at the temperature of 300 K, 600 K, 900 K, 1200 K and 1500 K, respectively. The structure of T′-Ca$_2$P monolayer can survive up to 1200 K and collapse with the temperature rises to 1500 K, indicating that the structure of T′-Ca$_2$P is likely thermally stable around 1200 K.

For the electronic properties, the band structure shows AFM-0D T′-Ca$_2$P is antiferromagnetic semiconductor with an indirect band gap as 0.25 eV (PBE) and 0.56 eV (HSE06), while those of other states (NM, FM and AFM of T and T′ phases) are all metal (as shown in Fig. 1e and Fig. S4† in ESI†). The comparison of the total DOS and partial DOS (as shown in Fig. 1f) of AFM-0D of T′-Ca$_2$P shows that the contribution of the Ca and P atoms is few at fermi level, indicating that the states around E$_f$ are mainly contributed by extra electrons located on the surfaces away from Ca atoms. The PDOS reveal that orbital contribution around E$_f$ is dominated by the Ca-$d$, higher than that of from Ca-$p$, Ca-$s$ and P-$p$. Compared to FM T-Ca$_2$P, the total atomic contribution slightly decreases, and the sharp DOS peak located around the fermi level splits, and shifts away from E$_f$ for both in AFM-0D and FM T′-Ca$_2$P, indicating there is directional Peierls transition in isotropic T-Ca$_2$P and result in anisotropic T′-Ca$_2$P.

For electride materials, the intriguing confined anionic electrons are always noticed. Electron localization function (ELF[31]) shows that the phase transition of Ca$_2$P leading a change for electron gas from isotropic 2DEG to anisotropic 0DEG, indicating the reconstructed interaction between adjacent Ca atoms in T and T′-Ca$_2$P within different magnetic states (see part V for more details of formation and distribution of electron gas in ESI†). The electron localization functions (ELFs) show the electron gas distribution in different magnetic states of T-$Ca_2P$ and T′-$Ca_2P$. The NM T-$Ca_2P$ shows the similar delocalized 2DEG to eletrene $Ca_2N$[30]. For FM T-$Ca_2P$, part of electrons located on the Ca-Ca bond move to the center of triangle Ca atoms, forming semi-1DEG. Unlike the isotropic T-$Ca_2P$, the electrons of the anisotropic FM T'-$Ca_2P$ are linearly distributed on the folds, forming a parallel 1DEG on both sides. In the AFM state, the electrons are most localized as 0DEG. In addition, by calculating the total energies of these systems, the energy of them gradually decreases as the dimension of electron gas decreases. Note that, the definition of nDEG (n=0, 1, 2, 3) is not quite clear so far, which are usually decided by the shape of the electron enrichment area away from atomic core shown in ELFs. To further understand the dimension change of electron gas as well as the phase stability of T-$Ca_2P$ and T'-$Ca_2P$, the average Crystal orbital Hamilton population (COHP)[43] of Ca-Ca interaction are calculated. As shown in Fig. 2b, the negative values of NM and FM of T-$Ca_2P$, and T'-$Ca_2P$ at the fermi level, imply different degree of anti-bonding occupation. However, the negative value of the T'-$Ca_2P$ phase is less than that of FM-T-$Ca_2P$ and NM-T-$Ca_2P$, indicating the T'-$Ca_2P$ is more stable than the T-$Ca_2P$ in bonding. The AFM-0D of T'-$Ca_2P$ phase has no anti-bonding states at the fermi level, leading to the most stability in bonding. The COHP analysis agreed well with the energy sequence of E(NM of T-$Ca_2P$) > E(FM of T-$Ca_2P$) > E(FM of T'-$Ca_2P$) > E(AFM-0D of T'-$Ca_2P$). Therefore, the stabilization through phases transition of $Ca_2P$ monolayer system results in the multiple changes of dimension of electron gas on its surfaces.

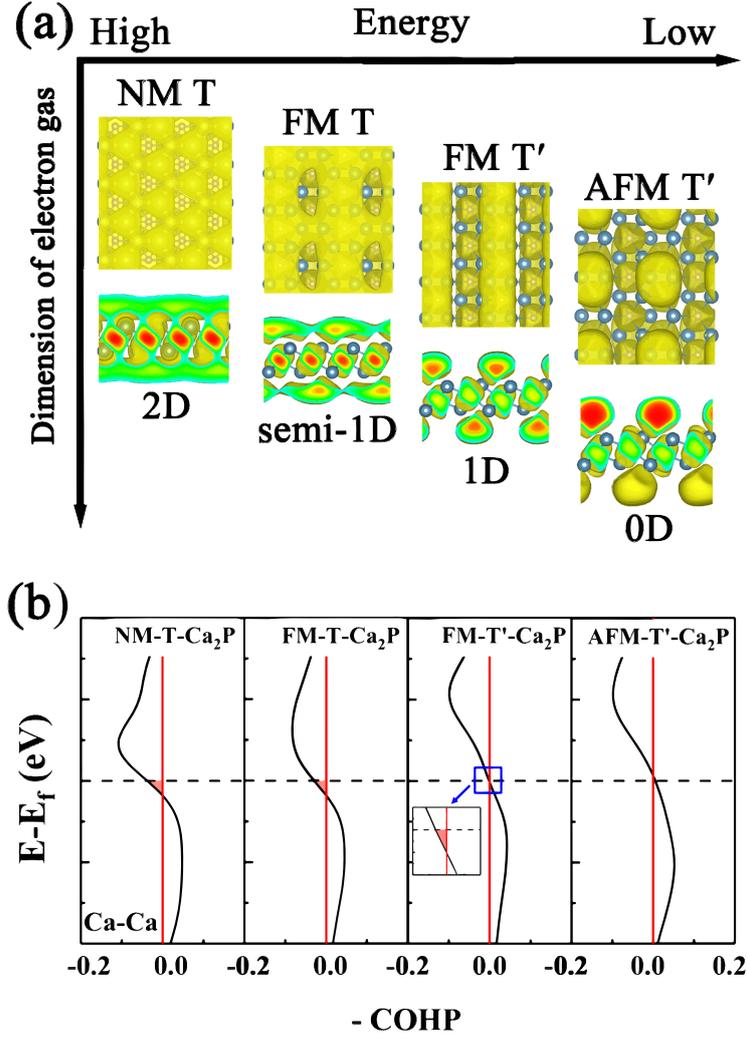

**Fig. 2** (a) The electron localization functions (ELFs) of NM-T-Ca$_2$P, FM-T-Ca$_2$P, FM-T′-Ca$_2$P and AFM-T′-Ca$_2$P. The isovalue is set as 0.20 e/Bohr$^3$. (b) Corresponding calculated average Crystal Orbital Hamilton Population (-COHP).

For mechanical properties, the elastic constants are calculated to be $C_{11}$ = 32.34 N/m, $C_{12}$ = 8.89 N/m, $C_{22}$ = 35.52 N/m, $C_{66}$ = 9.15 N/m, and them match the Born-Huang criterion[44] ($C_{11}, C_{22}, C_{66} > 0, C_{11}C_{22} - C_{12}^2 > 0$), implying that the T′-Ca$_2$P monolayer is mechanically stable. The in-plane Young's stiffness ($Y_{2D}$) basis of elastic constants are calculated to be $Y_{11}$ = 33.11 N/m and $Y_{22}$ = 33.08 N/m by Equation S1 (see part VI in ESI†). This indicated T′-Ca$_2$P is softer than Ca$_2$N ($C_{11}$ = $C_{22}$ = 61.54 N/m, $C_{12}$ =

17.27 N/m, $C_{66}$ = 148.02 N/m and $Y_{2D}$ = 56.69 N/m). And the $Y_{2D}$ of both of them is much lower than $MoS_2$[45](123 N/m) and graphene[46] (340 N/m).

To further investigate the interlayer interactions on the structure of T′-$Ca_2P$, the calculated cleavage energy ($E_{Cl}$) are shown in Fig. S7†, which is defined as $E_{Cl}$ = $[E_{tot}(m) - E_{tot}(m-1) - E_{tot}(1)]/S$, where m and S represents the number of layers and square of T′-Ca2P monolayer, respectively. The average $E_{Cl}$ as -0.044 eV/ Å$^2$ is twice that of graphite[47] (-0.021 eV/Å$^2$) but less than electride $Ca_2N$[30] (-0.067 eV/ Å$^2$), suggesting that the T′-$Ca_2P$ monolayer is less stable than graphene but favorable than $Ca_2N$ monolayer.

### 3.2    Na/K atom adsorption and migration on the T′-$Ca_2P$ monolayer surface

To evaluate the capability of metal atom adsorption on T′-$Ca_2P$ monolayer as an anode material for rechargeable battery, various adsorption sites and migration paths of a single Na/K atom are examined. A $2 \times 3 \times 1$ supercell of T′-$Ca_2P$ (including 24 Ca atoms and 12 P atoms) is constructed, which is large enough to avoid the interactions between neighboring Na/K atoms. Three stable adsorption sites on T′-$Ca_2P$ are identified (shown in Fig. 3a, the optimized structure as shown in Fig. S8†), of which the corresponding adsorption energies ($E_{Ads}$, vs. Na/K bulk) were calculated by Equation S2† (see part VII in ESI†). A comparison of the $E_{Ads}$ calculated by the PBE-GGA and DFT-D2 methods are shown in Fig. 3b. The $E_{Ads}$ obtained by the PBE-GGA shows that S1 site is the most favorable adsorption site for Na atom. In the case of K, S3 site is more stable than another adsorption site. However, with the DFT-D2 methods, the adsorption energies for all sites are slightly lower than the PBE-GGA results, while the S3 adsorption site is identified to be more favorable than the other sites for both cases of Na and K. In particular, the adsorption of K is stronger than Na at each site on T′-$Ca_2P$ surface. The charge density difference for both Na and K at the most favorable sites is calculated by Equation S3† and shown in Fig. S9a† and S9b†, of which the charges are accumulated between the Na/K and surface of T′-$Ca_2P$ monolayer. Bader charge[48] analysis shows that the Na and K atoms transfer 0.448 e and 0.592 e to T′-$Ca_2P$, respectively. Besides, based on the results of DOS of the configurations with Na and K adsorption, the T′-$Ca_2P$ monolayer gained a few extra electrons are still spin-polarized (as shown in Fig. S9c† and S9d†).

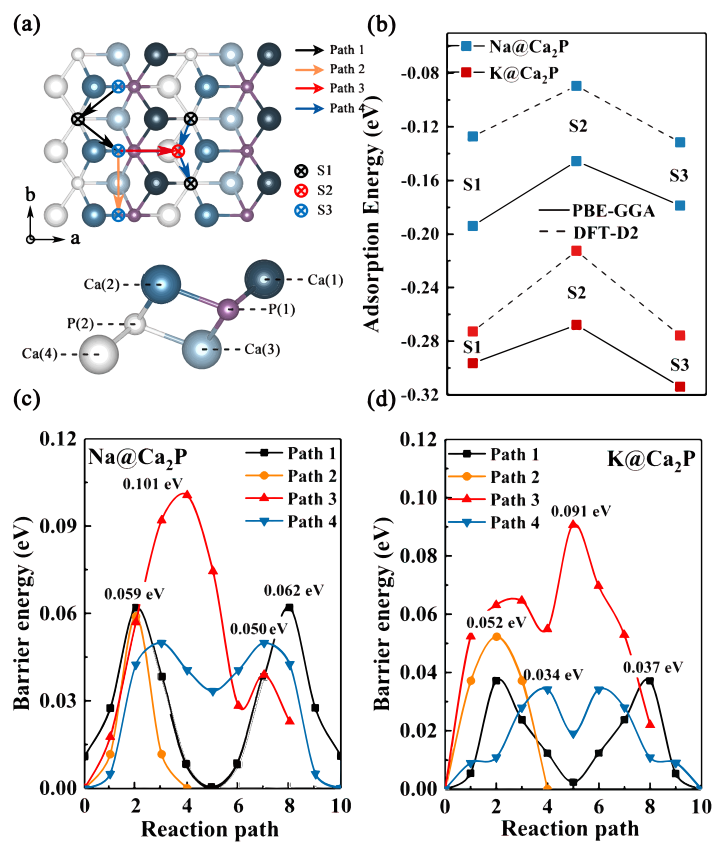

**Fig. 3** (a) Top and side view of the T′-Ca$_2$P monolayer, which the color points and lines represent adsorption sites and migration paths of Na/K on T′-Ca$_2$P monolayer, respectively. (b) Calculated adsorption energy for Na/K on the distinct adsorbing sites with PBE-GGA (solid lines) and DFT-D2 (dashed lines). The calculated energy barrier for Na (c) and K (d) on T′-Ca$_2$P monolayer.

The charge and discharge rate capability highly depend on the ion conduction of the metal atoms generally. As the results of the structural symmetry and adsorption stability, four possible pathways are considered. Specifically, the four paths correspond to: from S3 across S1 to S3 (Path 1), from S3 directly to S3 (Path 2), from S3 to S2 (Path 3) and from S1 cross S2 to S1 (Path 4) as shown in Fig. 3a and S10†. Note that some migration paths are neglected due to the predictable high energy barriers when the Na/K atoms cross the top site of Ca(1) and Ca(2). CI-NEB method[49] is used to calculated the energy barrier of Na and K migration along the four paths and the results (PBE-GGA) is plotted in Fig. 3c and 3d. For Na on the T′-Ca$_2$P, Path 4 (in direction *a*) and Path 3 (in direction *b*) are optimal migration pathways across the entire T′-Ca$_2$P surface, of which the energy barrier is 0.050 eV and 0.101 eV, respectively. With the DFT-D2 method, the Path 4 is still the optimal migration path, and the energy barrier is 0.040 eV. Particularly, the energy barrier value of Path 3 in direction *b* is lower than some that of reported anode materials for SIBs, such as graphene[50] (0.13 eV) and MoS$_2$[51] (0.11 eV), comparable to other electrenes materials, such as Ca$_2$N[31] (0.08 eV), Ba$_2$N[34] (0.012 eV) and Y$_2$C[32] (0.01 eV). In the case of K, Path 3 appear to be the more favorable pathway along direction *a* with a low energy barrier of 0.091 eV. In the direction *b*, the migration barrier for Path 4 is the lowest, which is about 0.034 eV. However, considering the great probability of K adsorbed at the S3 site, Path 1 should be the most favorable pathway. And the calculated migration barrier of K along Path 1 on T′-Ca$_2$P (0.037 eV for both PBE-GGA and DFT-D2) are also quite small compared with other PIBs anode materials, such as graphene[50] (0.10 eV), MoS$_2$[52] (0.063 eV), but slightly higher than that of other electrenes, such as Sr$_2$P[34] (0.016 eV) and Ba$_2$As[34] (0.01 eV). Zero-point energy (ZPE) and quantum mechanical tunneling (QMT) effects[53] are considered and evaluated for Na/K migration kinetics on T′-Ca$_2$P monolayer. The calculated energy difference between ZPE-corrected barrier ($E_b + \delta E_{ZPE}$) and classical barrier ($E_b$) are only as 1.13 meV. The tunneling effect is even much lower than the zero-point-energy and can be neglected (see part VII for more details of the calculations in ESI†). The results for migration energy barriers suggest fast migration of Na and K atoms on T′-Ca$_2$P surface due to the low energy barriers.

**3.3   Storage capacity and average open-circuit voltage of the T′-Ca$_2$P monolayer**

For practical application, the storage capacity and average open-circuit voltage of the batteries are both the key indicators for the electrode materials. Here, the $2 \times 3 \times 1$ supercell is still used for simulation. We assume the charge and discharge processes follow the typical half-cell reaction as follows:

$$Ca_2P + xNa^+ + xe^- \leftrightarrow Na_xCa_2P$$

$$Ca_2P + xK^+ + xe^- \leftrightarrow K_xCa_2P$$

For actual charging process, the affection of the pre-existing intercalated Na or K atoms is needed to be considered. Hence, the step adsorption energy ($E_{step}$ vs. Na/K bulk) are calculated by Equation S4† (see part IX in ESI†) to identify the stability of Na and K adsorption process. For Na adsorbed on T′-Ca$_2$P monolayer, we have calculated the adsorption numbers of 6, 12, 18, and 24 atoms on the monolayer. The fully relaxed structures of Na$_x$Ca$_2$P are presented in Fig. S11a†, of which the calculated $E_{step}$ are -0.226, -0.077, -0.155 and -0.281 eV with PBE-GGA (-0.244, -0.040, -0.177 and 0.310 eV with DFT-D2), respectively. However, the $E_{step}$ becomes a positive value as the second layers are adsorbed, suggesting that the saturated adsorption capacity of T′-Ca$_2$P within $2 \times 3 \times 1$ supercell is 24 Na atoms, and the corresponding chemical stoichiometry is Na$_2$Ca$_2$P. Similarly, for K on T′-Ca$_2$P, the calculated $E_{step}$

with PBE-GGA method are -0.267, -0.137, -0.201 and -0.013 eV (-0.270, -0.148, -0.218 and -0.089 eV with DFT-D2) for $K_{0.5}Ca_2P$, $K_1Ca_2P$, $K_{1.5}Ca_2P$ and $K_2Ca_2P$, respectively (as shown in Fig. S11b†). And the $E_{step}$ also becomes positive when more K atoms are adsorbed. The stoichiometric ratio under the saturation would be $K_2Ca_2P$. Therefore, the theoretical specific capacity of T′-$Ca_2P$ as anode for both SIBs and PIBs is calculated to be 482 mAh/g by Equation S5†. This value is higher than graphene (65 mAh/g for SIBs[54] and 279 mAh/g for PIBs[55]) and $MoS_2$ (335 mAh/g for SIBs[51] and PIBs[52]). Compared with other electrenes, the value of capacity for SIBs is lower than $Ca_2N$[31] (1138 mAh/g for SIBs) and $Y_2C$[32] (564 mAh/g for SIBs), and that for PIBs is higher than $Ba_2N$[34] (185 mAh/g for SIBs), $Ba_2P$[34] (175 mAh/g for PIBs), $Sr_2P$[34] (260 mAh/g for PIBs) and $Sr_2N$[34] (283 mAh/g for PIBs).

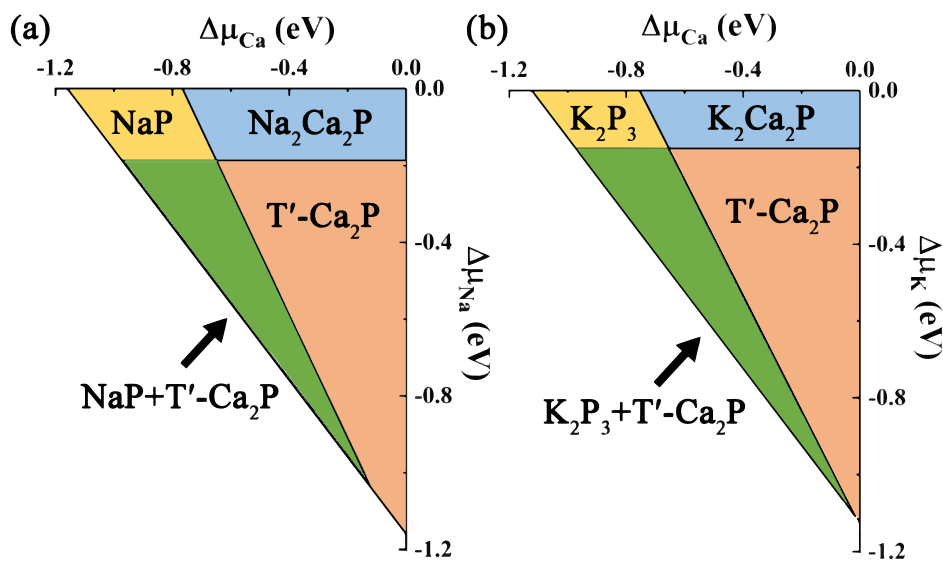

**Fig. 4** Thermodynamic phase diagrams of sodiated (a) and potassiated (b) T′-$Ca_2P$ monolayer.

To study the role of phase transition in thermodynamic stability of sodiated and potassiated T′-Ca$_2$P monolayer, the thermodynamic phase diagrams[56-59] of M$_2$Ca$_2$P (M = Na/K) are calculated, as shown in Fig. 4a and 4b (see part X for details of phase-diagram calculations in ESI†). There is a stable phase region for the formation of both Na$_2$Ca$_2$P and K$_2$Ca$_2$P in the allowed range of chemical potentials. In addition, varying the chemical potential from the P-rich to Na-rich/K-rich condition lead to the phase transition from T′-Ca$_2$P to Na$_2$Ca$_2$P/K$_2$Ca$_2$P, in particular the NaP/K$_2$P$_3$ and mix phase of NaP/K$_2$P$_3$ and T′-Ca$_2$P will appear at Ca-pool condition. The results mean that the Na$_2$Ca$_2$P and K$_2$Ca$_2$P phase are more facile to obtained in the T′-Ca$_2$P in the discharge process. Then, the structure stability of M$_2$Ca$_2$P (M = Na/K) are confirmed by 2D structure global search prediction carried out by using CALYPSO[60-61] code, that both of them are global minima in energy (see part X for details of CALYPSO calculations in ESI†).

To further investigated the dynamic reaction between Na/K atoms and T′-Ca$_2$P monolayer, AIMD simulations of the saturated adsorption structure M$_2$Ca$_2$P (M=Na,K) are computed at the temperature of 300 K and 500 K within the time of 5 ps, respectively. Fig. S13† shows the snapshots of the structure before and after performing AIMD simulations. There is no obvious structural destruction of Ca$_2$P to form M$_x$P$_y$ (M=Na,K) for the monolayer after adsorption of Na and K atoms. The AIMD simulation and thermodynamic phase diagrams both reveal that the T′-Ca$_2$P monolayer has good stability during the sodium and potassium insertion processes.

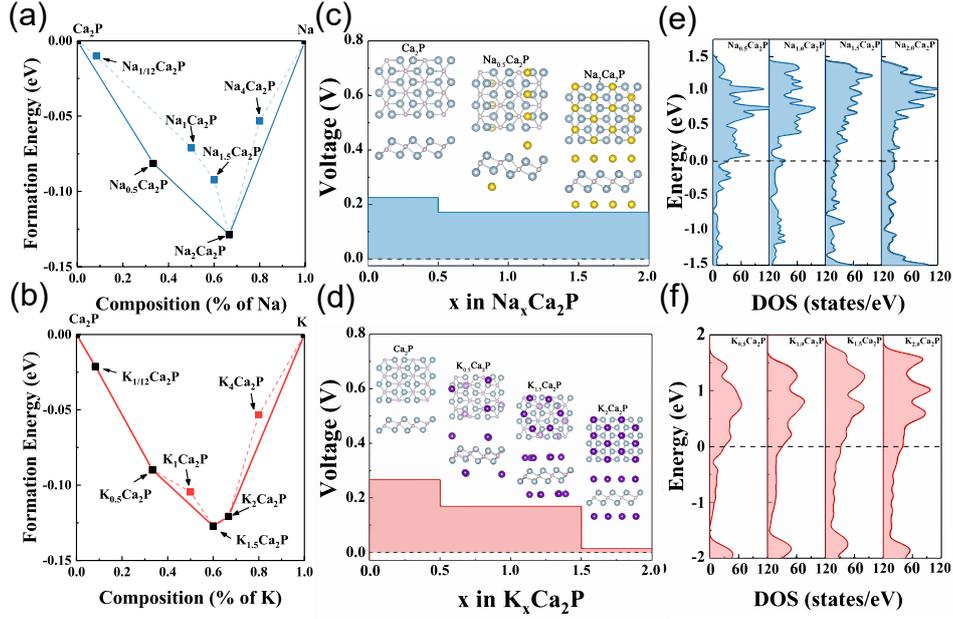

**Fig. 5** The calculated formation energy of $Na_xCa_2P$ (a) and $K_xCa_2P$ (b) (x = 0, 1/12, 0.5, 1, 1.5, 2 and 4), Na (c) and K (d) intercalation potential profile for $Na_xCa_2P$ and $K_xCa_2P$, respectively. The inset graph represents the thermodynamically stable structure of $Na_xCa_2P$ and $K_xCa_2P$. Total DOS of $Na_xCa_2P$ (e) and $K_xCa_2P$ (f) (x = 0.5, 1, 1.5 and 2), respectively. The Fermi levels are set to zero, the black dashed line represents the Fermi level.

Finally, to explore the most thermodynamically stable structure with different concentration of $M_xCa_2P$ (M = Na/K), the variation of formation energy convex hulls with adsorption concentration is plotted (as shown in Fig. 5a and 5b, calculated by Equation S6†, see part XI in ESI†). For Na adsorption, only the calculation points of $x$ = 0.5 and 2.0 are located on the convex hull, suggest the $Na_{0.5}Ca_2P$ and $Na_{2.0}Ca_2P$ are thermodynamically stable, the calculated average $V_{OC}$ decreases from 0.226 V to 0.171 V with PBE-GGA and from 0.244 V to 0.176 V with DFT-D2 (by Equation S7†, see part XI in ESI†), as shown in Fig. 5c. For K, the stable intermediate phases of $K_xCa_2P$ are with $x$ of 1/12, 0.5, 1.5 and 2.0. The corresponding average $V_{OC}$ curve decreases from 0.267 V to 0.013 V with PBE-GGA and from 0.270 V to 0.089 V with DFT-D2, as shown in Fig. 5d. It is noted that the small and positive $V_{Ave}$ value ensures the safety and efficiency during charge and discharge process.

Besides, the DOS of each intermediate phases of $M_xCa_2P$ (M = Na/K) are calculated and plotted in Fig. 5e and 5f. All the systems with different adsorption concentration of Na/K are metallic and their magnetic property decreases significantly and finally disappear when more metal atoms are adsorbed. The geometries after adsorption of Na and K atoms and the percentage change in the lattice constants of the system are checked and calculated to be in range of -2%~2.4% (See part XII for more details in ESI†). The ELFs of $Na_2Ca_2P$ (Fig. S14a†) and $K_2Ca_2P$ (Fig. S14b†) shows that, the electron gas between alkali metal layers and $Ca_2P$ monolayer indicated a stable adsorption.

## 4. Conclusions

In conclusion, we proposed and evaluated the potential of an electrene T′-$Ca_2P$ monolayer reconstructed from unstable T-$Ca_2P$ as anode material of SIBs/PIBs by first-principles simulations. By the calculated phonon spectra and AIMD simulation, the thermal and dynamical stabilities of T′-$Ca_2P$ monolayer are confirmed. Impressively, the calculated electronic structure indicates that the T'-$Ca_2P$ is a magnetic semiconductor electrene, of which the ground state is antiferromagnetic with 0D electron gas as well as 0D spin density on its both surfaces. The suitable single atom adsorption and anisotropic but low migration energy barriers of 0.050/0.101 eV (for Na) and 0.037/0.091 eV (for K) on T′-$Ca_2P$ surface implicates the excellent rate performance of T′-$Ca_2P$. The specific capacity for both Na and K was verified up to 482 mAh/g, and the $V_{OC}$ value of Na and K is predicted to be in a range of 0.171-0.226 V and 0.013-0.267 V, respectively. With all of these extraordinary characteristics, the T′-$Ca_2P$ is expected to be a new promising anode material for SIBs and PIBs with excellent electrochemical performance and potential applications.

## Author Contributions

†These authors contributed equally.

## Conflicts of interest

There are no conflicts to declare.

† **Electronic supplementary information (ESI): See DOI: 10.1039/x0xx00000x**

## Acknowledgment

This work was supported by the Anhui Provincial Natural Science Foundation (No. 2008085QA33), the National Natural Science Foundation of China (No. 11775261) and the Hefei Advanced Computing Center## References